\input{epsf}

\documentclass[preprint,preprintnumbers,amsmath,amssymb]{revtex4}

\usepackage{graphicx}

\begin{document}
\draft
Published as:  Chan, M .H. and Ehrlich, R., Astrophys. and Space Sci., 349, (1), 407- 413, (2014).
\\ 
\title{Sterile neutrino fits to dark matter mass profiles in the Milky Way and in galaxy clusters}


\author{Man Ho Chan}
\address{The Chinese University of Hong Kong,  Shatin, New Territories, Hong Kong}
\email{mhchan@phy.cuhk.edu.hk}
\author{ Robert Ehrlich}
\address{George Mason University, Fairfax, VA 22030}
\email{rehrlich@gmu.edu}



\begin{abstract}
In recent papers it was claimed that SN 1987A data supports the existence of 4.0 eV and 21.4 eV active neutrino mass eigenstates, and it was suggested that such large active neutrino masses could be made consistent with existing constraints including neutrino oscillation data and upper limits on the neutrino flavor state masses.  The requirement was that there exist a pair of sterile neutrino mass states nearly degenerate with the active ones, plus a third active-sterile doublet that is tachyonic ($m^2 <0$).   Here, independent evidence is presented for the existence of sterile neutrinos with the previously claimed masses based on fits to the dark matter distributions in the Milky Way galaxy and four clusters of galaxies.  The fits are in excellent agreement with observations within the uncertainties of the masses.  In addition, sterile neutrinos having the suggested masses address the "cusp" problem and the missing satellites problem, as well as that of the "top down" scenario of structure formation -- previously a chief drawback of HDM particles.   Nevertheless, due to the highly controversial nature of the claim, and the need for two free parameters in the dark matter fits, additional confirming evidence will be required before it can be considered proven.
\end{abstract}

\maketitle
\section{Introduction}
In several recent papers evidence was presented based on an analysis of SN 1987A for the existence of two active neutrino mass eigenstates having masses $4.0 \pm 0.5 eV$ and $21.4 \pm 1.2 eV$, hereafter referred to as the "claim" \citep{Ehrlich1, Ehrlich2}.  The present paper reviews that claim and presents new independent evidence supporting it.  It is shown that the problem of dark matter within galaxies and galaxy clusters can be resolved using a pair of sterile neutrinos having the very same masses as claimed previously.  Even though the claim was based on active neutrinos from SN 1987A, their existence would require sterile neutrinos having nearly the same masses, so as to be compatible with the very small values of $\Delta m_{atm}^2$ and $\Delta m_{sol}^2$ measured in neutrino oscillation experiments.  Of course, in such a case the small $\Delta m^2$ values observed to date in most oscillation experiments would need to be between an active and a sterile mass state, contrary to what is normally assumed.  While the main new result presented here involves the dark matter fits, we first review the prior work supporting the "claim," and discuss how such large neutrino masses could be made consistent with other observations.

\section{Prior work making the claim}

The values of the neutrino masses have been a matter of considerable interest both theoretically and experimentally ever since they were shown to be nonzero by virtue of the existence of neutrino oscillations.  It is of course only the mass differences between pairs of the mass eigenstates, i.e., $\Delta m^2$, that are revealed in neutrino oscillation experiments not the masses themselves, with the relation between the flavor and mass eigenstates and their respective masses being through the mixing matrix $U_{ij}$ which relates the mass and flavor state masses:  

\begin{equation}
(a) \hspace{0.25in} \nu_{Fi}=\Sigma U_{ij} \nu_{Mj} \hspace{0.5in} (b) \hspace{0.25in} m^2_{Fi}=\Sigma |U_{ij}|^2 m^2_{Mj}
\end{equation}

Empirically, for the flavor state masses only upper limits exist from various decay processes, such as tritum beta decay, which as of 2013 yields an upper limit on the electron neutrino mass $m<2.0 eV$ \citep{PDG}.   However, given the manner in which the effective flavor state masses are found (Eq. 1), the existence of a third mass state that is tachyonic ($m^2<0$), can easily result in arbitrarily small flavor state masses (for a suitable choice of $U_{ij}$) even if the first two mass states have values as large as those claimed here.  Thus, there need be no conflict with the low mass limit for the electron neutrino mass or those of other flavors, so that a conflict with the sum of the flavor state masses from cosmology (around $\Sigma m < 0.27 eV$) can be similarly avoided.\citep{PDG}  

The ability of an experiment to measure the mass for a mass eigenstate is severely limited for eV-scale masses, and perhaps the only feasible sources for doing so are the supernovae with their long baseline and briefness of the initial pulse.   In conventional analyses of SN 1987A it is assumed that one can only obtain a limit on the electron neutrino mass based on the spread in neutrino arrival times for different energy neutrinos.  However, since electron neutrinos consist of a mixture of mass states that separate out en route to Earth (and are then detected as electron neutrinos), one could in principle measure the masses of the individual mass states provided the emissions at the source were confined to a very small time interval.  

Thus, the only departure our SN 1987A analysis makes from a conventional one is that it is assumed that the varying neutrino arrival times are primarily a reflection of their travel times rather than the times of their emission.   The recent pair of papers making the claim of 4.0 eV and 21.4 eV masses was based on an analysis of SN 1987A,\citep{Ehrlich1, Ehrlich2} and it relies on an idea that was first investigated much earlier\citep{Huzita, Cowsik}.  The recent papers, however, extended the earlier analyses by reducing the uncertainties in the claimed masses, and showing a way such large neutrino masses could be compatible with what is empirically known about neutrino mass limits \citep{Ehrlich1,Ehrlich2}.  

As noted, the claim rests on the assumption that most of the 24 neutrinos observed from SN 1987A were emitted during a time much less than the 15 second burst itself, possibly in as short a time as 1 second.   If one simplistically assumes simultaneous neutrino emissions then the individual neutrino arrival times at the detectors $t$ can be expressed in terms of the light travel time from the supernova $t_0$ and that neutrino's mass $m$ and energy $E.$  Based on relativistic kinematics we have: $1-m^2c^2/E^2=v^2/c^2 =t_0^2/(t_0+t)^2\approx1-2t/t_0,$ so that

\begin{equation}
\frac{1}{E^2} = \frac{2t}{m^2t_0}
\end{equation}

Note that by Eq.~(2), $t=0$ is the time that infinite energy ($v = c$) neutrinos would have arrived at the detector -- which given the neutrino time distribution in the detectors, is probably within $\pm 0.5 s$ the same as setting $t=0$ for the first arriving neutrino in each of the three unsynchronized detectors \citep{Ehrlich1}.  Note that Eq.~(2) implies that on a plot of $1/E^2$ versus $t$ all neutrinos having a fixed mass $m$ lie on a straight line of slope $M=2/(m^2t_0)$.  Thus, if neutrinos having two distinct masses were present among the 24 they would all lie close to one of two straight lines having slopes $M_1=2/(m_1^2t_0)$ and $M_2=2/(m_2^2t_0)$ that pass through the origin.  Remarkably, this pattern is exactly what the data show (within the uncertainties in $1/E^2$) for every one of the 24 observed neutrinos \citep{Huzita, Cowsik, Ehrlich1}.  

\subsection{Achieving consistency with other observations}

There are many reasons for skepticism about the claim, however we believe none is fatal.  One reason is that it is unlikely that all the 24 SN 1987A neutrinos were in fact emitted during a time as short as 1s, and some core collapse modelers calculate that much of the neutrino emission occurs over an extended time.\citep{Hudepohl}  Nevertheless, other modelers do show the emissions to be very concentrated in time \citep{Totani}.   For example, Totani et al. shows $73\%$ emitted in the first second and $82\%$ in the first 2 s (values based on a numerical integration of their Fig. 1).\citep{Totani}   Finally, we note that the goodness of the two straight line fit (from which the claim was inferred) is not significantly impaired if there is a limited spread in emission times.  Thus, in a Monte Carlo simulation in which 24 events are generated, assuming an emission time distribution and energy spectrum following those of Totani\citep{Totani}, and  assuming the masses of individual neutrinos are as claimed, the agreement of the fit to two straight lines is often just as good as for the actual data. 

Achieving consistency between such large claimed masses and existing constraints on the neutrino masses requires an unconventional model of 3 active and 3 sterile neutrino mass states.  In this 3 + 3 model there are three nearly degenerate active-sterile doublets, as depicted in Fig. 1 of Ehrlich(2013)\citep{Ehrlich2}.  The masses of the first two active-sterile doublets are $m_1=4.0 eV$ and $m_2=21.4 eV$ and the third active-sterile doublet mass is tachyonic, and has a conjectured mass: $m_3^2=-0.2 keV^2.$  

The basis of this conjectured mass for the tachyonic doublet follows from the observed values from oscillation experiments for $\Delta m_{sol}^2$ and $\Delta m_{atm}^2$ plus the $\Delta m^2\approx 1 eV^2$ suggested by short baseline reactor neutrino experiments\citep{Mention}.   If one assumes $\Delta m_{sol}^2$ and $\Delta m_{atm}^2$ represent the splittings of the two tardyonic doublets, and $\Delta m^2\approx 1 eV^2$ is that for the tachyonic doublet, then assigning $m_3^2=-0.2 keV^2,$ one then finds virtually identical fractional mass splitting of each of the three doublets, i.e., $\frac{\Delta m_1^2}{m_1^2}=\frac{\Delta m_2^2}{m_2^2}=\frac{\Delta m_3^2}{m_3^2}=5.0\times 10^{-6}$.  Further suggestive evidence for the value of tachyonic mass is discussed in Ehrlich(2013)\citep{Ehrlich2} based on the limited amount of published Kamiokande data for the one hour preceding the main burst.  

We have already noted how this 3 + 3 model could be consistent with upper limits on the flavor state masses, and neutrino oscillation data.  There is also the matter achieving consistency with various cosmological constraints.  For example the large value of the tachyonic mass $m_3^2=-0.2 keV^2$ might seem to be at variance with the stringent upper limits on any tachyonic neutrino masses that might be the source of dark energy, e. g., that set by Davies, $|m|<0.33 eV.$\citep{Davies}  However, dark energy depends on the average energy density throughout space, and hence it depends not on the mass state masses but the sum of the flavor state masses which are produced in particle reactions in the early universe.  Thus, a mass as large as $m_3^2=-0.2 keV^2$ would of necessity be a very small component of the flavor states (to make their masses all close to zero), and hence the large value of this mass would be offset by its very small spatial density in its contribution to dark energy.   

How does the existence of large sterile neutrino masses comport with other cosmological constraints?   One recent analysis of a variety of cosmological probes including the cosmic microwave background (WMAP7+SPT), Hubble constant (HST), galaxy power spectrum (SDSS-DR7), and supernova distances (SDSS and Union2 compilations) shows that the sterile neutrino model provides a good fit to each of the considered data sets, and no single probe manages to decisively disfavor the sterile neutrino model, although there is a problem in achieving consistency with all the data sets.  In summary, Joudaki et. al conclude that it is premature to either rule out the existence of several massive sterile neutrinos or claim this model is cosmologically preferred.\citep{Joudaki}  In fact, recent data fits from Cosmic Microwave Background (CMB) do 
not rule out the possibility of 100 percent HDM. \citet{Angus} has shown 
that the CMB spectrum can still be fitted well for all dark matter to be 
$\sim 
10$ eV sterile neutrinos. Besides, the free streaming length of neutrinos 
is given by $l \sim 2(m/1~\rm eV)^{-1}$ Mpc \cite{Petraki}. For $m_s=4$ eV 
and $m_s=21.4$ eV, the free streaming lengths are 0.5 Mpc and 90 
kpc respectively, which are smaller than the corresponding halo sizes for 
a typical cluster ($\sim 3$ Mpc) and a typical galaxy ($\sim 100$ kpc). 
Therefore, sterile neutrinos with these masses can form structures within 
galaxies and clusters.

Clearly, given its highly controversial nature the claim of $4.0 eV$ and $21.4 eV$ neutrinos needs to be supported by other observations before it will be taken seriously.   In fact, it is more impressive if such a confirmation comes from an entirely different area from supernova neutrinos.  Therefore it is particularly interesting that the analysis reported in the present paper on the implications of sterile neutrinos for the dark matter in galaxies and galaxy clusters does in fact provide that independent sort of evidence.  

Normally, one assumes that if neutrinos are a dark matter candidate it would require that the mass states be sterile rather than active, primarily because the active neutrino masses are believed to be too small to cause the needed "clumping."  However, this is not the case for the 3 + 3 model in which the there are three nearly degenerate active and sterile states having sizable masses.  Moreover, while we may we may have (perhaps unwisely) continued to refer to the dark matter neutrinos as sterile in this paper it should be clear that both the active and sterile neutrinos having a given mass (either $m_1=4.0 eV$ and $m_2=21.4 eV$) can contribute to the neutrino halo.  Nevertheless, the proportion of active and sterile mass states to the dark matter halos makes no difference in the model since the active-sterile masses are nearly identical.  

Note that unlike dark energy which depends on the average energy density throughout all space, dark matter tends to be concentrated in bound systems such as galaxies and clusters, and its distribution is sensitive to the masses of the neutrino mass eigenstates, and not the flavor eigenstates.  That is because, even though neutrinos are produced in reactions in specific flavor states the admixture of the mass states making up the flavor states separate out the longer the neutrinos propagate.  In fact the tachyonic component will not be gravitationally bound to a galaxy or cluster, leaving only the $m_1$ and $m_2$ states to contribute to their dark matter halos.

\section{Sterile neutrinos as hot dark matter in galaxies and clusters}

If they exist most of the sterile neutrinos can be produced through active-sterile neutrino
oscillation by resonant or non-resonant mechanisms during the early epoch of the Big Bang \citep{Dodelson,Shi}.  Traditionally, the dark matter problem has been mainly addressed by the assumption it is cold.  However, recent literature indicates that cold dark matter cannot explain the existence of flat cores at the center of clusters (the "cusp" problem)\citep{Tyson,Sand,deBlok,Newman} and the relative absence of satellite galaxies\citep{Spergel}. The existence of warm or hot dark matter better reconciles theory and the observations by solving these problems.  In fact, simulations show that a flat core can be produced and the number of dwarf galaxies surrounding the Milky Way would be smaller if the dark matter particles are warmer \citep{Chan,Cho}.   Furthermore, since the formation time of a dark matter halo varies inversely with the mass of the particles and since the sterile neutrino mass used is larger in the galaxies than in clusters, the formation time of the galactic halo will be shorter than for clusters. Hence, in this scenario, the galaxies would completely form before the clusters, which matches the theorized bottom-up formation scenario in the structure formation.    If sterile neutrinos are the dark matter particles, their halo profile should match the dark matter profile in both clusters and galaxies. 

\subsection {Derivation of sterile neutrino dark matter mass profile}

In this section, we derive the sterile neutrino halo profile and compare it to the observed mass profile in clusters and the Milky Way.    

The pressure of the sterile neutrinos is given by
\begin{equation}
P= \frac{4 \pi g_s}{3m_sh^3} \int_0^{\infty}
\frac{p^4dp}{e^{\frac{\epsilon- \mu}{kT_s}}+1},   
\end{equation}
where $m_s$, $g_s$, $T_s$, $\epsilon=p^2/2m_s$, $\mu$ and $p$ are the 
rest mass, degrees of freedom, temperature, kinetic energy, chemical 
potential and the
momentum of the sterile neutrinos respectively. If $m_s \sim 1$ eV for a 
$10^{15}M_{\odot}$ cluster mass and $m_s \sim 20$ eV for a 
$10^{11}M_{\odot}$ galactic mass, the number 
densities are about $10^5$ cm$^{-3}$ and $10^6$ cm$^{-3}$ respectively. 
For $T_s \sim 1$ K, this number density is nearly the same order of 
magnitude 
of the quantum number density $n_Q=(2m_skT_s/h^2)^{3/2} 
\sim 10^6-10^7$ cm$^{-3}$. Therefore, most sterile neutrinos are in the 
classical regime or they are only 
slightly degenerate ($e^{\frac{\epsilon- \mu}{kT_s}} \gg 1$). Then, Eq.~(3) can be written as
\begin{equation}
P= \frac{4 \pi g_s}{3m_sh^3} \int_0^{\infty}e^{- \frac{\epsilon-
\mu}{kT_s}}(1-e^{- \frac{\epsilon- \mu}{kT_s}})p^4dp.
\end{equation}
Let $u=\sqrt{\epsilon/kT_s}=p/\sqrt{2m_skT_s}$ and
$v=\sqrt{2 \epsilon/kT_s}=p/\sqrt{m_skT_s}$, Eq.~(4) becomes
\begin{equation}
P= \frac{4 \pi g_s}{3m_sh^3} \left[(2m_skT_s)^{5/2}e^{\mu/kT_s}
\int_0^{\infty}e^{-u^2}u^4du-(m_skT_s)^{5/2}e^{2 \mu/kT_s}
\int_0^{\infty}e^{-v^2}v^4dv \right].
\end{equation}  
The integrals in Eq.~(5) are both equal to $3 \sqrt{\pi}/8$.
Therefore we get
\begin{equation}
P= \frac{\pi^{3/2}
g_s}{2m_sh^3}e^{\mu/kT_s}(m_skT_s)^{5/2}(2^{5/2}-e^{\mu/kT_s}).
\end{equation}  
Additionally, the number density of the sterile neutrinos is given by
\begin{equation}
n= \frac{4 \pi g_s}{m_sh^3} \int_0^{\infty}
\frac{p^2dp}{e^{\frac{\epsilon- \mu}{kT_s}}+1}.
\end{equation}  
By using the same approximation method, we get
\begin{equation}
n= \frac{\pi^{3/2}g_s}{h^3}e^{\mu/kT_s}(m_skT_s)^{3/2}(2^{3/2}-e^{\mu/kT_s}).
\end{equation}  
Therefore we have
\begin{equation}
P= \frac{nkT_s}{2} \left(\frac{2^{5/2}-e^{\mu/kT_s}}{2^{3/2}-e^{\mu/kT_s}} \right).
\end{equation}  
In the classical limit $e^{\mu/kT_s} \rightarrow 0$, Eq.~(9)
reduces to the ideal gas law $P=nkT_s$.   However, if we expand 
$e^{\mu/kT_s}$ up to first order to allow
a slight degeneracy, we have
\begin{equation}
P=nkT_s(1+2^{-5/2}e^{\mu/kT_s}).
\end{equation}  
So that from Eq.~(8), we then have
\begin{equation}
e^{\mu/kT_s} \approx \frac{2^{-3/2}nh^3}{\pi^{3/2}g_s}(m_skT_s)^{-3/2}.
\end{equation}
Substituting Eq.~(11) into Eq.~(10), and writing $n$ as $n(r)$ to reflect its dependence on radial distance, we obtain for the final equation of state for the slightly degenerate neutrinos
\begin{equation}
P(r)=n(r)kT_s+ \frac{n^2(r)h^3}{16 \pi^{3/2}g_s}(m_skT_s)^{-3/2}(kT_s)^{-1/2}.
\end{equation}
By using the hydrostatic equilibrium condition, since $\rho(r)=m_sn(r)$ we find 
\begin{equation}
\frac{dP(r)}{dr}=-\frac{Gm_sM(r)n(r)}{r^2},
\end{equation}
and finally, the overall mass profile $M(r)$ can be found as the sum of the dark matter profile and the observable baryonic mass profile $M_B(r)$
\begin{equation}
M(r)=M_D(r)+M_B(r)=\int_0^r4 \pi r^2 m_s n(r)dr+M_B(r),
\end{equation}
The three coupled eqs.~(10-12) cannot be solved in closed form, but a numerical solution is possible if one assumes values for the parameters $m_s$, $T$ and $\rho_0$ and then iterates in radial steps outwards from $r=0.$

This leaves only the issue of the function $M_B(r)$.  In clusters, the baryonic mass profile is commonly expressed in the $\beta$-model profile \citep{King}
\begin{equation}
M_B(r)=\int_0^r4 \pi r^2 \rho_{B0} \left(1+ \frac{r^2}{r_c^2} \right)^{-3 
\beta/2}dr,
\end{equation}
where $\rho_{B0}$, $r_c$ and $\beta$ are the parameters that can be fitted by the surface brightness profile of a cluster. In the Milky Way, the baryonic mass is mainly composed by the 
bulge and disk. These components can be modelled by the gravitational potential model \citep{Flynn}
\begin{equation}
M_B(r)=- \frac{r^2}{G} \left[ \frac{d}{dr} \left( \sum_{n=0}^2 
\frac{M_{cn}}{\sqrt{r^2+r_{cn}^2}}+ \sum_{n=0}^3 
\frac{M_{dn}}{\sqrt{r^2+(c_n+b)^2}} \right) \right], 
\end{equation}
where $M_{cn}$, $M_{dn}$, $r_{cn}$, $c_n$ and $b$ are fitted 
parameters in 
the model \citep{Flynn}.
Since the sterile neutrino temperature is nearly homogeneous, just like 
the cosmic microwave background, we can assume that the sterile neutrino 
is isothermal in an entire cluster and in the Milky Way.  

Note that the larger the sterile neutrino mass, the smaller will be the 
halo size. Therefore it is assumed that the dark matter profile in the 
Milky Way is mainly characterized by the $21.4 \pm 1.2$ eV sterile 
neutrinos and that of galaxy clusters by the $4.0 \pm 0.5$ eV sterile 
neutrinos.  In Fig.~1-3, we fit the derived total mass profiles 
from 
Eqs.~10-14 and the NFW mass profile from CDM model \cite{Navarro} together 
with the data from four large clusters (A478, 
A1413, A2204 and PKS0745) \citep{Pointeco} and the rotation curve of the 
Milky Way \citep{Sofue1,Sofue2}, respectively. We have chosen these four 
particular clusters from Ref.[23] because the outermost data points are 
about 1 Mpc from the center so that dark matter is dominated at that 
region. The 
central density of the sterile neutrino halo $\rho_{s0}$ and the 
sterile neutrino temperature $T_s$ are the only free parameters adjusted in the fits. The neutrino temperatures used in Figs.~1 and 2 are $T_s=0.1-0.5$ K (for $m_s=4.0$ eV) and $T_s 
= 0.07$ K (for $m_s=21.4$ eV) respectively. The parameters used in 
the baryonic mass profile from each cluster can be obtained from 
observations \citep{Pointecouteau,Pratt,Vikhlinin,Hoshino,Sanders,Reiprich,Chen}. 
As can be seen, the fits from our model agree very well with the 
observed data. The reduced $\chi^2$ in cluster and Milky Way fits in 
our model are $2.84-6.37$ and $1.27$ respectively. Compare with the CDM 
model, the corresponding reduced $\chi^2$ are equally good: $0.95-8.32$ and $1.43$. 
Therefore, they agree equally well with the observational data. However, 
as mentioned above, the CDM model suffers from some major defects such as 
core-cusp problem and missing satellite problem while sterile neutrinos 
do not. It suggests that 
$4.0$ eV sterile neutrinos may be in 
fact the major component of dark matter in galaxy clusters within 1.5 Mpc, 
and that $21.4$ eV sterile neutrinos may be its main component in the Milky Way within 25 kpc.

\section{Discussion}

In summary, here we have found excellent agreement with observations when using the claimed masses of two sterile neutrinos, $4.0 \pm 0.5 eV$ and $21.4 \pm 1.2 eV$ that had been inferred previously from an analysis of SN 1987A.   Although such good fits (all within the cited $1 \sigma$ errors on the masses) provides confirmation for the claimed masses existence, the highly controversial nature of the claim and the use of two free parameters to achieve the fits means that additional observations will be needed to support the claim.   The most direct confirmation might involve a neutrino oscillation experiment sensitive to $\Delta m^2=21.4^2 - 4.0^2= 442 eV^2$.  For example, at a 30 MeV neutrino energy we find an oscillation wavelength of about 0.167 m, which could be readily observed, particularly using a high intensity source with a strong (30\%) monochromatic component, such as the Oak Ridge Spallation Neutron Source (SNS) \citep{Efremenko}.   If we are fortunate enough to have another supernova in our galaxy any time soon, the good sensitivity of Super-Kamiokande and other detectors also could give such a clear confirmation or refutation of the claim.  In fact, Super-K is even sensitive enough to search for a supernova burst from Andromeda (2 events expected).  Surprisingly, even the observation of only 2-3 events associated with Andromeda could determine whether the claim of two mass eigenstates is correct, in light of the pulse spreading with source distance required by Eq.~(2).  However, it is important that such a search would need to be conducted differently from the one actually employed in a 2007 Super-K supernova search by Ikeda et al.\citep{Ikeda}, as described in a recent paper \citep{Ehrlich3}.

\begin{figure}
\includegraphics[width=140mm]{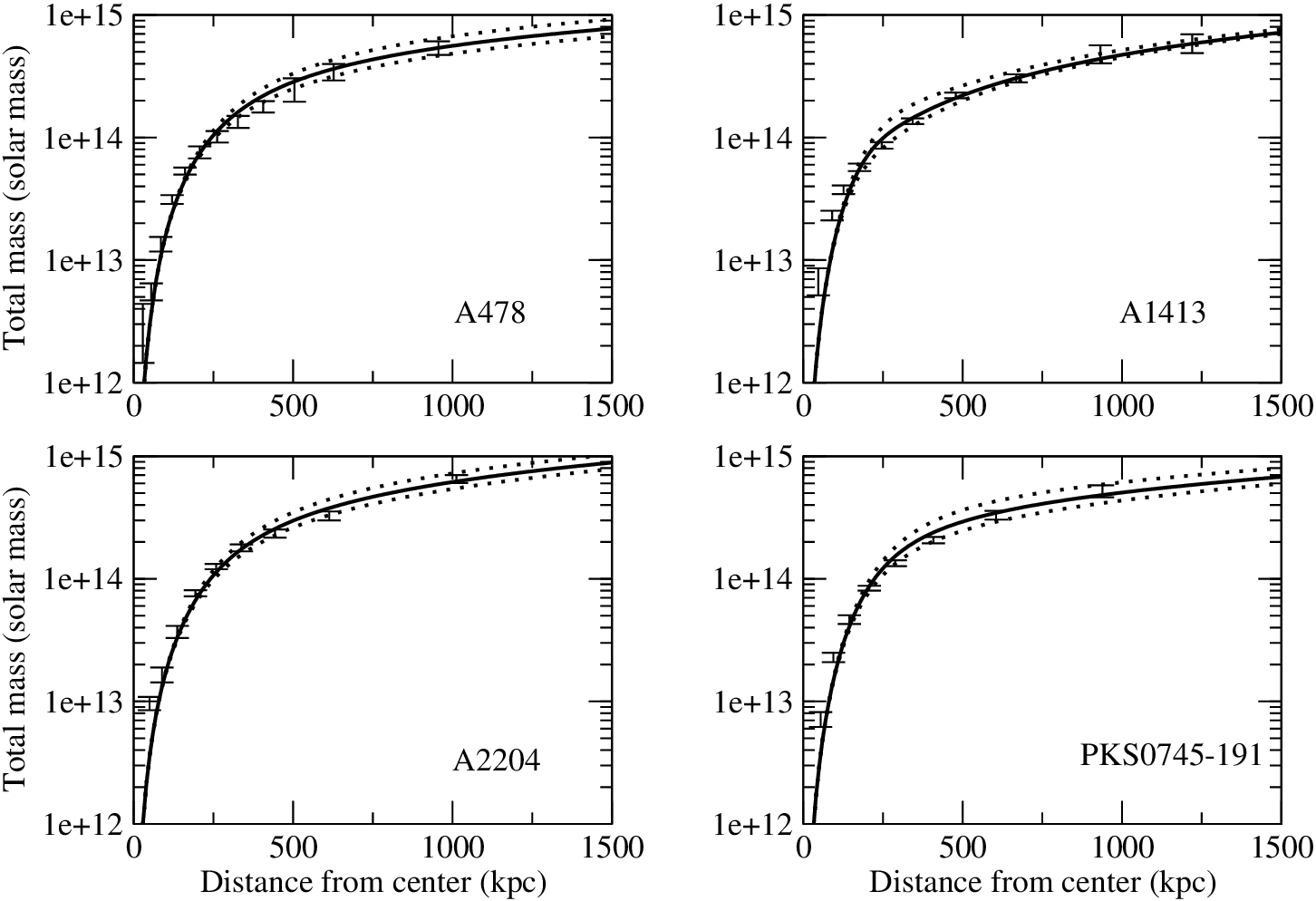}
 \caption{Observed and fitted total mass profiles for clusters A478, A1413, A2204 and 
PKS0745-191. Given the $1 \sigma$ errors on $m_1=4.0\pm 0.5 eV$, the sterile neutrino masses used in the fits are 3.5 eV (dotted line), 4 
eV (solid line) and 4.5 eV (dotted line) in each cluster. The central 
sterile neutrino densities, the sterile neutrino temperatures and the 
reduced $\chi^2$ values for the four 
clusters are: $(\rho_{s0},T_s,\chi_{\rm reduced}^2)$ are ($1.4 \times 
10^{-25}$ g cm$^{-3}$, 0.5 K, 2.84) (A478), 
($1.5 \times 10^{-25}$ g cm$^{-3}$, 0.1 K, 6.37) (A1413), ($1.5 \times 
10^{-25}$ g 
cm$^{-3}$, 0.5 K, 4.84) (A2204) and ($2.2 \times 10^{-25}$ g cm$^{-3}$, 
0.4 K, 5.48) (PKS0745-191) 
respectively.}
\end{figure}

\begin{figure}
\includegraphics[width=140mm]{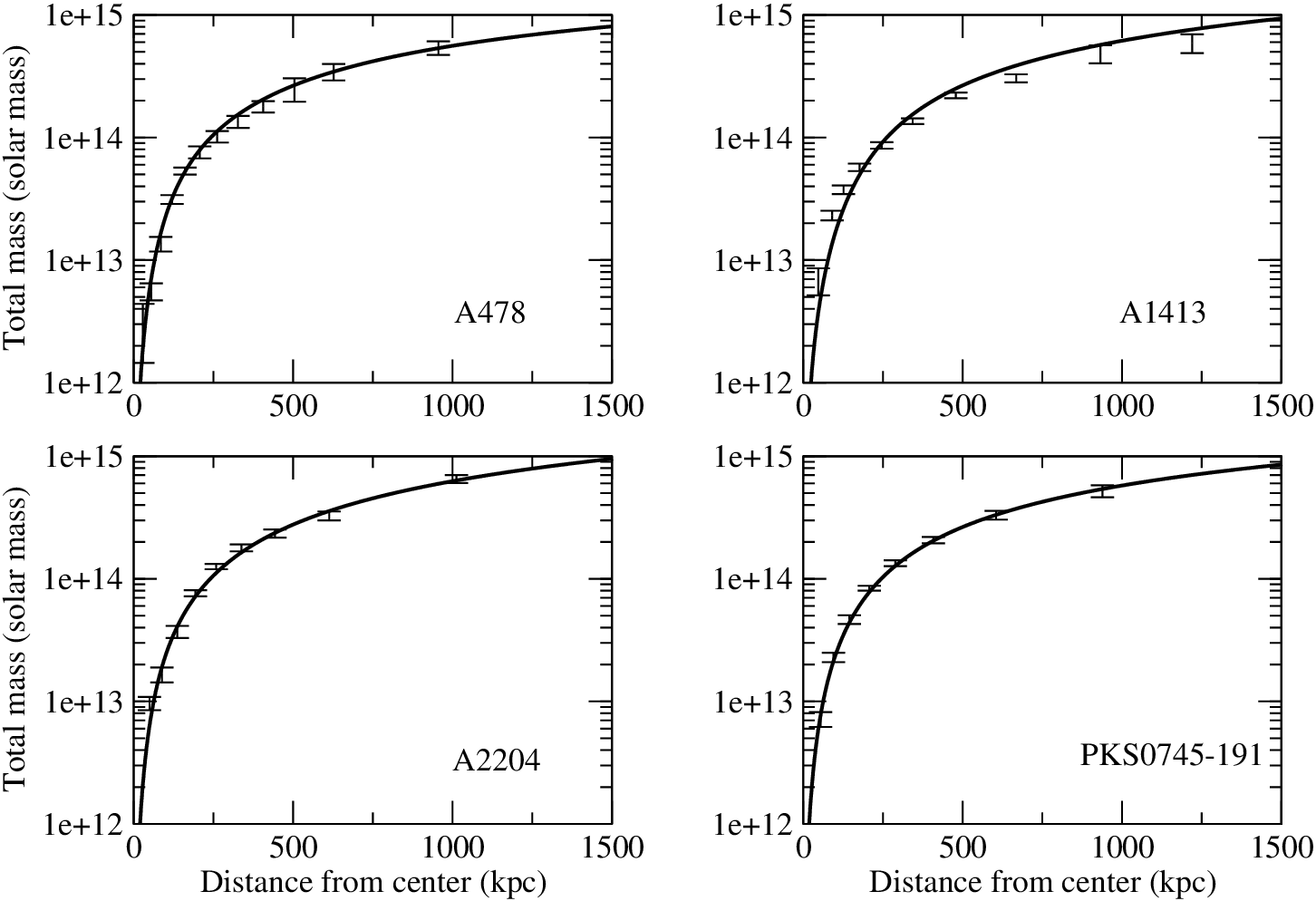}
 \caption{The fitted total mass profiles by CDM model for clusters A478, 
A1413, A2204 and PKS0745-191. The reduced $\chi^2$ values for the four 
clusters are 0.95, 8.04, 2.32 and 1.28 respectively.}
\end{figure}

\begin{figure}
\includegraphics[width=140mm]{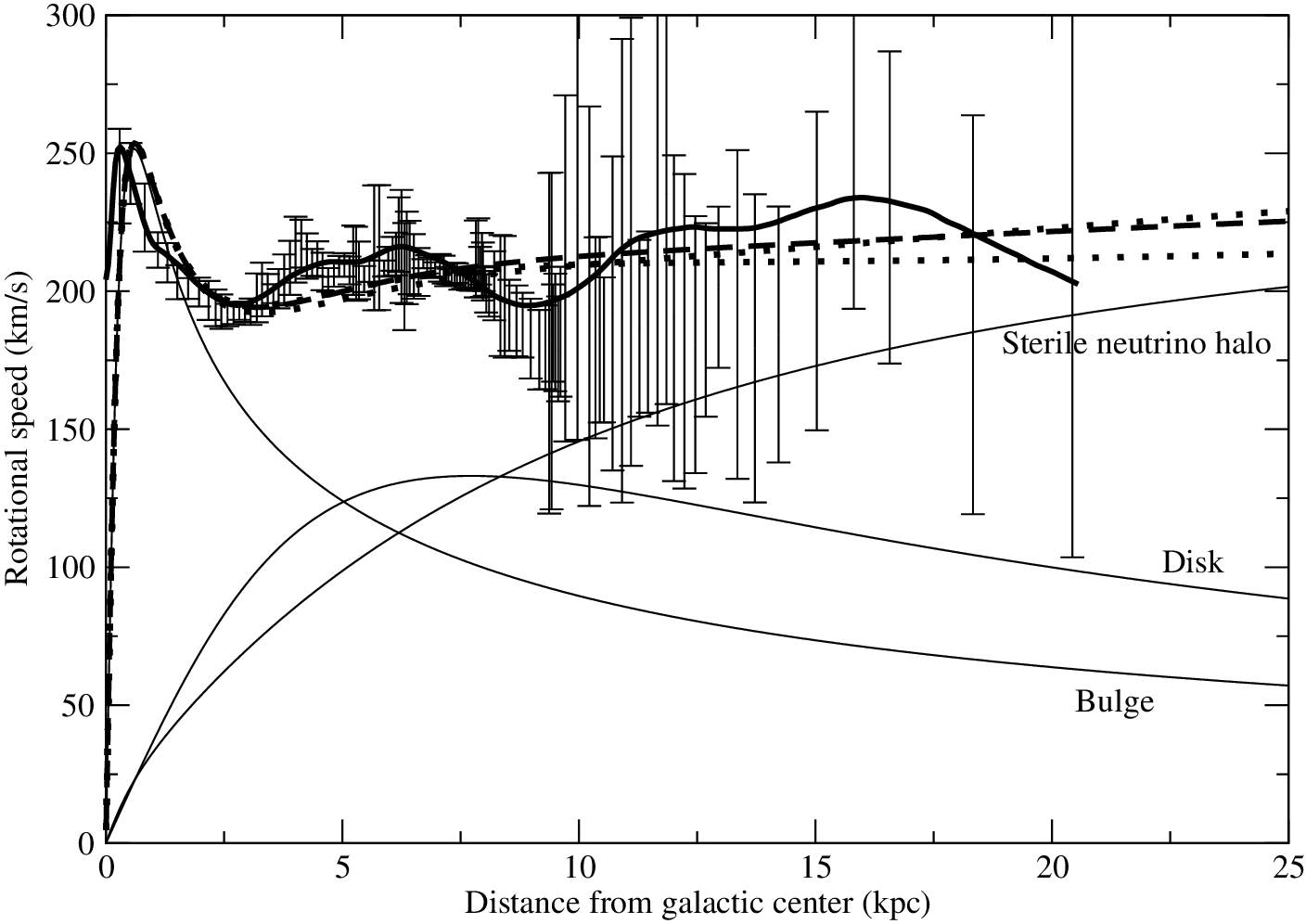}
 \caption{Observed and fitted rotation curve of the Milky Way. The solid dark curve is the observed data 
obtained by Sofue et al.(1999) and the error bars are the latest data 
obtained by Sofue(2011). The components of the sterile neutrino halo, disk and bulge that contribute to the total rotation curve are shown. The dashed line and the dotted 
lines are the rotation curves fitted by using $m_s=21.4$ eV (central density $=7.3 
\times 10^{-24}$ g cm$^{-3}$) and $m_s=20.2(22.6)$ eV,  (central density $= 7.3
\times 10^{-24}$ g cm$^{-3}$ for $m_s=21.4$ eV and central density =$6
\times 10^{-24}$ g cm$^{-3}$ and $8.5 \times 10^{-24}$ g cm$^{-3}$ for
$m_s=20.2$ eV and 22.6 eV respectively). The sterile neutrino temperature 
is 0.07 K. The dash-dotted line is the fits obtained by using CDM model 
\cite{Navarro,Sofue2}.}
\end{figure}

\end{document}